\documentclass[useAMS,usenatbib]{mn2e}

\usepackage{graphicx}
\usepackage{latexsym}
\usepackage[hypertex]{hyperref}

\begin{document}

\def\p{\partial}
\def\f{\varphi}
\def\t{\tilde}
\def\ds{\displaystyle}

\title[Dissipationless discs]{Accretion and plasma outflow from dissipationless discs.}

\author[S.V. Bogovalov and S.R. Kelner]{S.V. Bogovalov${}^1$, S.R. Kelner${}^{1,2}$\\
 ${}^1$Moscow Engineering Physics Institute (State University), Moscow, Russia\\
${} ^2$Max-Planck-Institut f\"ur Kernphysik, Heidelberg, Germany }
\date{ }
\pubyear{2007}

\maketitle

\begin{abstract}
We consider an extreme case of disc accretion onto a gravitating centre when the viscosity in the disc is negligible. The angular momentum and the rotational energy of the accreted matter is carried out by a magnetized wind outflowing from the disc. The outflow of matter from the disc occurs due to the Blandford \& Payne(1982) centrifugal mechanism. The disc is assumed to be cold. Accretion and outflow are connected by the conservation of the energy, mass and the angular momentum. The basic properties of the outflow, angular momentum flux and energy flux per particle in the wind, do not depend on the details of the structure of the accretion disc. In the case of selfsimilar accretion/outflow, the dependence of the rate of accretion $\dot M$ in the disc depends on the disc radius $r$ on the law $\dot M \sim r^{{1\over2(\alpha^2-1)}}$, where $\alpha$ is a dimensionless Alfvenic radius. In the case of $\alpha \gg 1$, the accretion in the disc is provided by very weak matter outflow from the disc and  the outflow predominantly occurs from the very central part of the disc. The solution obtained in the work provides mechanism which transforms the gravitational energy of the accreted matter into the energy of the outflowing wind with efficiency close to $100\%$. The final velocity can essentially exceed Kepler velocity at the site of the wind launch. This mechanism allows us to understand the nature of the astrophysical objects with low luminosity discs and energetic jet-like outflows.
 \end{abstract}

\section{Introduction}

Conventional theory of accretion discs proposed by  \cite{shakurasunyaev} was successful in interpretation of observations of the accretion discs in the binary systems. This theory satisfactory predicts the general properties of the discs around compact objects. Nevertheless some important phenomena connected with the accretion appeared incompatible with the conventional theory. The most important among them are the accretion discs with anomalously low luminosity and jet-like outflows from the objects. The best example in this regard is AGN M87.

Advection dominated disc model (ADAF model ) was proposed to explain the under luminous discs, in particular disc around BH located in the Galactic Centre \citep{adafs,narayan}. This theory, however does not solve the problem of jet-like outflow especially in the cases when the kinetic luminosity of jets is comparable (like in the case of SS433 case) or even exceeds ( like in the case of the jet from M87) the bolometric luminosity of the object. In fact, these problems
including low luminosity accretion discs, very efficient outflow from the discs and its collimation into jets could be internally connected.

Observations show that jets are strictly connected with the disc accretion. In all jet detections a signature of a disc accretion was found as well.
It is important for understanding of the mechanism of the jet ejection that this phenomena is not connected with the specific nature of the central object. Jets are formed irrespectively to the fact that the central object is black hole ( like in the case of jet from AGNs), neutron stars or protostar as it takes place in all cases of the jets from Young stellar objects. It is reasonable to assume that the jets are directly connected with the accretion mechanism itself rather than with the nature of the central object. However, in this case one observational fact needs to be explained. In all cases when it was possible to observe the base of the jets it appears that the jets are launched from the very central part of the disc which produce impression that central object could be connected some way with the process of the jet ejection. At least the fact that jet is ejected from the very central part of the accretion disc rather then from all the disc surface demands explanation.

Another difficult problem is that in some cases the mechanism of ejection is surprisingly efficient. For example, total bolometric luminosity of M87 does not exceed $10^{42} \rm ergs/s$ \citep{bolometry}, while the total kinetic luminosity of the jet from M87 is as high as $10^{44} \rm ergs/s$ \citep{lkin,lkin2}. Thus, if to estimate the gravitational energy release on the basis of conventional theory of \cite{shakurasunyaev}, the kinetic energy luminosity of the jet from M87 exceeds by two orders of magnitudes the gravitational energy released at the accretion. The conventional models of the disc accretion (Shakura \& Sunyaev type or ADAFs) do not predict the existence of such objects. The example with M87 jet shows that one needs to explore new regimes of accretion.

The most evident modification is incorporation into the model of the magnetic field.
It has already been widely recognized that the magnetic field has important impact and may play leading role in the ejection of the plasma from accretion discs. In this regard, two processes are of special interest. First one is related the idea of \cite{blandford}. They have demonstrated that the magnetic field results into instability of the particles at the Kepler orbit. If the angle between the force line of the magnetic field and the disc plane is less than $60^{\circ}$ the particles are freely ejected from the disc by centrifugal force. The second process was proposed by \cite{pelletier}. They argue that the winds from the accretion disc can carry out noticeable part of the angular momentum of the accreted material increasing the accretion rate. Together, these work clearly demonstrate that the magnetic field of the disc results into the outflow from the disc at rather general conditions and this wind can carry out essential part of the accreted material angular momentum.

In this work we consider an extreme case when the angular momentum of the disc is carried out by the wind. The viscous stresses are fully neglected. We solve the problem of the disc accretion under these conditions selfconsistently with the problem of the wind outflow from the disc. Fortunately, to provide selfconsistency of the processes of accretion and outflow one do not need to know information about the detailed structure of the accretion disc. The laws of conservation of mass, energy and angular momentum appeared sufficient to provide the selfconsistency of the processes of accretion and outflow.

The paper is organized as follows. In the first section we discuss the qualitative picture of the outflow due to the \cite{blandford} centrifugal instability and related structure of the poloidal magnetic field. In the second section the basic equations and connection of the accretion and outflow are discussed based on of the conservation laws.
In the third section the selfsimilar solutions of the problem are presented. In the last section we discuss the physical sense of the solution.

\section{Qualitative picture of the accretion under affect of wind.}

The idea that the wind can carry out essential fraction of the angular momentum from the disc has been explored in \citep{ferreira1,ferreira2}. The approach of Ferreira is analog of the pioneering studies by \cite{gena} of the disc accretion onto gravitating centre from a magnetized cloud surrounding the gravitating centre. They considered the accretion as the process of diffusion of the matter across the magnetic field lines rooted into interstellar medium. Actually, this kind of flow is similar to the Hartman flow in a channel \cite{landay8}. Taking into account that the plasma typically has very high electric conductivity it is necessary to assume in this scenario an existence of a mechanism which reduces the electric conductivity of plasma on a few orders of magnitude to provide efficient accretion. This scenario has another problem. The velocity of the plasma outflow from the disc exceeds the velocity of all MHD perturbations. Therefore no signal from the cloud can propagate to the disc along the field lines. Therefore, there is an internal inconsistency of the model. From one side it requires that the field lines are rooted into the cloud to prevent advection of the field lines with the matter. On the other hand the cloud can not affect on the magnetic field in the disc because it is causally disconnected from it.

\begin{figure}
\centerline{\includegraphics[width=0.4\textwidth,angle=0.0]{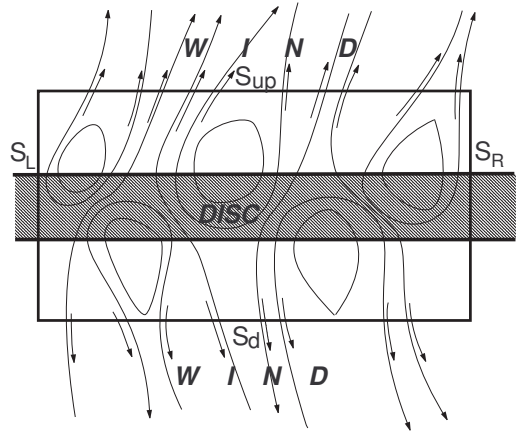}}
\caption{The structure of the magnetic field in the accretion disc and in the outflowing wind. The disc is shadowed. The magnetic field lines in the disc are distributed chaotically. At the base of the wind from the disc all the magnetic field lines are opened. Their direction is accidental as well. therefore the total magnetic flux leaving one side of the disc equals to zero. The box drown in black thick lines is the region of integrations of conservation laws connecting the properties of the disc and the wind. }
\label{scheme}
\end{figure}

Here we suggest another approach which does not face these kind of problems.
The source of the magnetic field in our model is the plasma of the accretion disc itself like Sun is the source of the magnetic field on the solar surface and in the interplanetary space. It is reasonable to expect that in the disc the magnetic field lines are distributed rather chaotically as suggested in \cite{blandford} and is shown in Fig.~\ref{scheme}.

In the physical model of Ferreira et al. the matter have to diffuse across the poloidal magnetic field lines. In our approach the matter falls down onto the gravitation centre together with the frozen-in magnetic field. Since dissipation processes are not necessary for this, the process of accretion and outflow can be considered in the limit of ideal magnetohydrodynamics where no heating of the matter. Therefore, the disc is cooled to low temperature even if the plasma was hot initially.

Thus, it is reasonably to expect that the disc is geometrically thin with thickness
\begin{equation}
{h\over r}={c_s\over V_k},
\end{equation}
where $c_s$ is the sound velocity in the disc, $V_k$ is the Kepler velocity of the disc and $r$ is the cylindrical radius.
The outflow of plasma from the cold disc occurs due to mechanism specified by Blandford \& Payne (1982). If the angle between the magnetic field line and the
plane of the disc is less than $60^{\circ}$ then a particle motion on the Kepler orbit becomes unstable. Any perturbation of the radius of the orbit results into sliding of the particle along the field line outward from the gravitating centre. Therefore, the outflow of the plasma from the accretion disc can occur even if the disc is cold.

\section{Equations.}

\subsection{Basic equations}
In this work we study a steady state axisymmetric flow of an ideal plasma. The condition of ideality has a form
\begin{equation}
{\bf E}+{1\over c}{\bf v\times B}=0,
\label{frozen}
\end{equation}
where $\bf E$ is the electric field, $\bf B$ - magnetic field and $\bf v$ is the velocity of the plasma.

It is convenient to consider MHD equations for ideal plasma in the form of conservation laws. According to \cite{landau2} the energy-momentum of plasma satisfies to the equation
\begin{equation}
{\p T_{ik}\over x_k}=0,
\label{tik}
\end{equation}
where $T_{ik}$ is the energy momentum tensor in the form
\begin{equation}
T_{ik} = \left\{
\begin{array}{lllll}
 T_{00} & q_1 & q_2 & q_3 \\
 q_1 & \sigma_{11} & \sigma_{12} &\sigma_{13} \\
 q_2 & \sigma_{21} & \sigma_{22} &\sigma_{23} \\
 q_3 & \sigma_{31} & \sigma_{32} &\sigma_{33} \\
\end{array}
\right\}
\end{equation}
Where $T_{00}=e+{B^2\over 8\pi}$ is the sum of the thermal and magnetic field energy densities. We consider here only nonrelativistic flows. Therefore the term in $T_{00}$ connected with the electric field is omitted. Spatial components of the tensor $\sigma_{ik}$ are as follows
\begin{equation}
\sigma_{ik}= \rho v_iv_k +p\delta_{ik}- {1\over 4\pi}(B_iB_k-{1\over 2} B^2 \delta_{ik}).
\end{equation}

Equation (\ref{tik}) still does not contain gravitational force.
To take into account the gravitational force it is sufficient to add into right hand part of Eq.(\ref{tik}) with a
spatial components of the tensor a term $-\rho GM {{\bf R}\over R^3}$ so that the equation for $\sigma_{ik}$ becomes as follows
\begin{equation}
{\p \sigma_{ik}\over\p x_k}=-\rho GM {R_i\over R^3}.
\label{sigmaik}
\end{equation}
Here $R_i$ is the component of radius-vector from the position of the gravitating centre.

The energy density flux $q_i$ is as follows
\begin{equation}
q_i=\rho v_i\left({v^2\over 2} -{GM\over R}\right)+ {c\over 4\pi}[E\times B]_i\,.
\end{equation}
In this expression the gravitational field is taken into account explicitly.
Therefore the equation for the energy density flux remains as follows
\begin{equation}
 {\p q_{k}\over\p x_k}=0
\end{equation}
for the steady state flow.

In the Cartesian  coordinates the angular momentum tensor is introduced as  $m_{ik}= \varepsilon_{imp}x_mT_{pk}$ \citep{landau2}, where
$\varepsilon_{imp}$ is unit antisymmetric tensor.
In this case the density flux of the $z$ component of the angular  momentum is as follows
\begin{equation}
l_i=\rho v_i rv_{\varphi}-{1\over 4\pi}rB_iB_{\varphi}.
\label{anlgle}
\end{equation}

The conservation equations for the matter and the magnetic fluxes are
\begin{equation}
{\p \rho v_{k}\over\p x_k}=0,
\label{mass}
\end{equation}
and
\begin{equation}
{\p B_{k}\over\p x_k}=0.
\label{bconserv}
\end{equation}

These equations are supplemented by the condition of stationarity of the magnetic field
\begin{equation}
{\rm curl}\,\bf E=0.
\label{rote}
\end{equation}
This condition means that ${\p {\bf B}\over \p t}=0$.

\subsection{The role of the induction equation in the problem of accretion.}

In the limit of axisymmetric flow an azimuthal component of Eq.(\ref{rote}) and frozen-in condition (\ref{frozen}) for the same component give a couple of equations
\begin{equation}
{\p\over \p r}(rE_{\f})=0,
\label{rote2}
\end{equation}
and
\begin{equation}
E_{\f}+{1\over c}\,(v_zB_r-v_rB_z)=0.
\label{frost}
\end{equation}
Here $r$ is the distance from the symmetry axis.

The solution of Eq.(\ref{rote2}) gives that
\begin{equation}
E_{\f}={A\over r},
\label{varphi}
\end{equation}
where $A$ is some constant. This solution diverges at $r\to0$. On this reason it is assumed that $E_{\f}=0$ \citep{mestel}.
This is a conventional assumption for study of axisymmetrical MHD winds.

\cite{iannis} was the first who pointed out that if we deal with the accretion of an ideal plasma onto a gravitating centre $E_{\f}$ is not equal to zero. Indeed, as it follows from Eq.(\ref{frost}), $E_{\f}$ at the equator is equal to
${1\over c}v_rB_z$ because $B_r \approx 0$ and $v_z=0$ at the centre of the disc. At accretion $v_r\ne 0$, therefore, $E_{\f} \ne 0$ as well. Thus, in the region of the accretion flow $ E_{\f}$ can not be neglected because it is connected directly with the radial velocity of the plasma in the disc.

\begin{figure}
\centerline{\includegraphics[width=0.33\textwidth,angle=0.0]{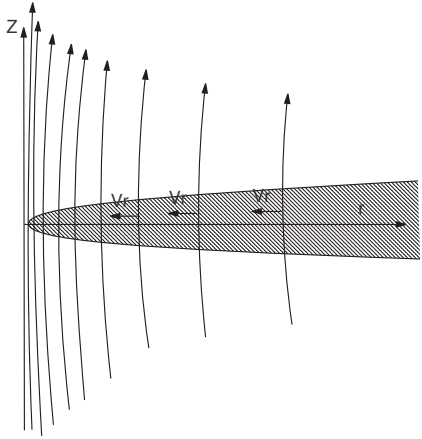}}
\caption{The advection of the magnetic field lines of one direction to the rotational axis results into infinite accumulation of the magnetic field flux. There is no way to annihilate the magnetic field with this topology.}
\label{problem1}
\end{figure}

\begin{figure}
\centerline{\includegraphics[width=0.33\textwidth,angle=0.0]{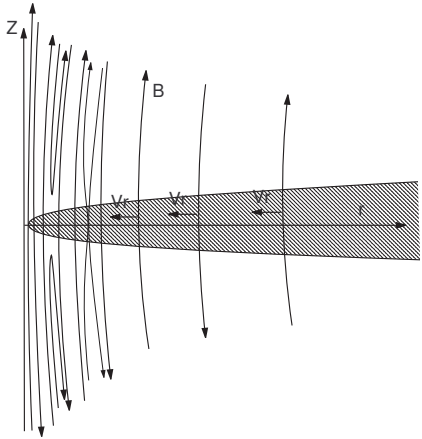}}
\caption{The advection of the magnetic field lines of different directions to the rotational axis does not result into infinite accumulation of the magnetic field flux because in average the magnetic flux on one side of the disc equals to zero and there is way to annihilate the field lines in the process of reconnection.}
\label{problem2}
\end{figure}

At first glance the advection of the magnetic field to the gravitating centre results into accumulation of the magnetic field lines at the centre and to infinite growth of the magnetic field. This happens if the direction of the magnetic field is the same at one side of the disc as it is shown in Fig.~\ref{problem1}. Actually this is unrealistic case. The direction of the magnetic field on one side of the disc varies accidentally so that in average the magnetic flux leaving one side of the disc is zero. That is why at the advection of this magnetic field to the gravitating centre there is no accumulation of the magnetic field flux because it equals to zero in average on the surface of the disc. The infinite growth of the magnetic field at the centre is also prevented by the reconnection of the magnetic field lines as it shown in Fig.~\ref{problem2}.

As was noted in \cite{BT}, the dynamics of an ideal plasma is invariant
with respect to the redirection of the magnetic field
lines. Therefore the dynamics of plasma in the fields shown in
Fig.~\ref{problem1} and Fig.~\ref{problem2} is the same.
With this in mind, we are able to consider an
outflow of plasma from the accretion disc with the azimuthally symmetrical magnetic field
shown schematically in Fig.~\ref{problem1}.

\subsection{Along field line MHD equations of the wind}

 We consider the case when $E_{\f} \ne 0$. As it follows from Eq.(\ref{frost}) in this case
the velocity has a poloidal component $v_{\perp}$ orthogonal to the poloidal magnetic field line.
Nevertheless, this component of the electric field can be neglected at the consideration of dynamics of the wind
 under the condition $v_{\perp} \ll v$, where $v$ is the full velocity
of the plasma. If to take into account that $v \sim V_k$ and that $v_{\perp} =c\,{E_{\f}\over B_p}$ the toroidal electric field can be neglected under the condition
\begin{equation}
{B_{p0} r_0^{}\over B_p r} {v_r\over V_k} \ll 1.
\label{cond1}
\end{equation}

It follows from this condition that the toroidal electric field can be neglected if the radial velocity in the disc $v_r \ll V_k$.
But this is not enough to provide the condition (\ref{cond1}). The product $B_p r$ should not drop down too strongly compared with the initial value. This means that if for example, poloidal field falls down as $r^{-2}$ there is limitation on the distance where the toroidal electric field can be neglected. Further we will be interested in the solution of the problem at the distance small compared with the size of the accretion disc. In this region $E_{\f}$ can be neglected. The conclusion that in the limit $v_r \ll V_k$ affect of the toroidal magnetic field on dynamics of the wind can be neglected has been shown explicitly for the selfsimilar solutions by \cite{kelner}.

\begin{figure}
\centerline{\includegraphics[width=0.35\textwidth,angle=0.0]{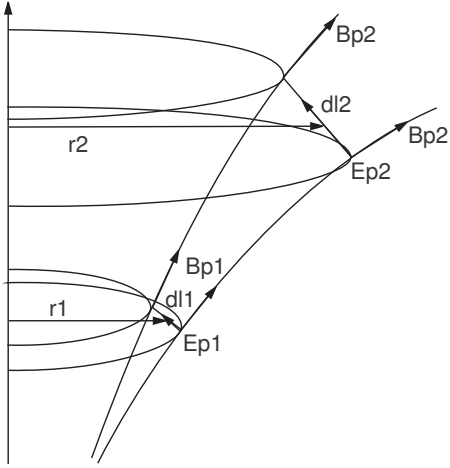}}
\caption{Fluxes of the magnetic field, matter, energy and angular momentum between any two close field lines with the cross section $dS=2\pi rdl$ are conserved. The condition ${\rm curl}\, \bf E=0 $ gives that the product $E_pdl$, where $dl$ is the distance between these field lines is also conserved.}
\label{bflux}
\end{figure}

In the region where $E_{\f} = 0 $ the poloidal velocity is directed along the poloidal magnetic field. This directly follows from Eq.(\ref{frost}). In this case we have
\begin{equation}
l_ p=\rho r v_p \left(v_{\f}-{B_p\over 4\pi \rho v_p}\,B_{\f} \right),
\end{equation}
for the angular momentum flux density along a poloidal filed line and
\begin{equation}
q_p=\rho v_p\left( {v^2\over 2} - {GM\over R} - \Omega r\, {B_p\over 4\pi \rho v_p}\,B_{\f} \right)
\end{equation}
for the energy density flux along a poloidal field line. If to take into account that the fluxes of the angular momentum $\sigma _ {\f p}dS$, energy $q_pdS$,
matter $r\rho v_p$ and magnetic field flux $B_p dS$ are conserved as it is demonstrated in Fig.~\ref{bflux}, it can be obtained that the following two integrals of motion take place along the field lines
\begin{equation}
 rv_{\f}-{1\over f} rB_{\f}= L,
\label{mom}
\end{equation}
where $f=4\pi \rho v_p/B_p$ and
\begin{equation}
 {v^2\over 2} - {GM\over R} - {1\over f}\,\Omega rB_{\f}=E.
\label{energy}
\end{equation}
The first equation from this couple is the conservation of the angular momentum per particle and the second one is the conservation of the energy per particle along a field line.

The frozen-in condition for the poloidal component of the electric field gives that
\begin{equation}
 r\Omega B_p + v_pB_{\f} = v_{\f} B_p.
\label{frosen2}
\end{equation}
We take into account in Eq.(\ref{frosen2}) that due to conservation of the magnetic field flux $B_pdS$ and product $E_pdl$, which is constant due to induction equation (\ref{rote}) there are relationships $E_r=-r\Omega B_z/c$ and $E_z=r\Omega B_r/c$.
Combining Eq.(\ref{frosen2}) with Eq.(\ref{mom}) gives that
\begin{equation}
rv_{\f}= {Lfv_p-r^2\Omega B_p\over fv_p-B_p}.
\end{equation}
Denominator of this expression goes to zero at the Alfvenic point where $v_p=B_p/\sqrt{ 4\pi \rho}$. The nominator of this expression must equal to 0 in this point to provide regularity of $v_{\f}$. From this condition we obtain that the momentum per particle equals to $L=\Omega\,r_A^2$, where $r_A$ is the radius at the Alfvenic point.

\subsection{The connection of the accretion and outflow}

Irrespective to the inner structure of the disc, the disc and the wind are connected by the conservation laws.
Let us consider a fragment of the disc as it is shown in Fig.~\ref{scheme}. A~rectangular region is shown by thick black lines. The upper and lower boundaries of this region are located at the base of the wind from the disc. All the magnetic field lines of the wind are rooted here.

Conservation laws in integral form are
\begin{equation}
\oint_S l_k\,dS_k=0,
\end{equation}
for the angular momentum, and
\begin{equation}
\oint_S \rho v_k\,dS_k=0
\end{equation}
for the mass conservation.
Integration here is performed along a closed surface surrounding the volume.

Let us perform integration of the angular momentum flux over a surface which includes a small part of the disc shown in Fig.~\ref{scheme}.
The surface consists of the sides $S_L$ at radius $r_1$, side $S_R$ at radius $r_2$ and upper and down sides $S_{up}$ and $S_d$.
Integration gives
\begin{equation}
\begin{array}{l}
\ds \left.-\int_{-h/2}^{h/2} r^2\rho v_{\f} v_r\,dz\right|_{r1} +\left.\int_{-h/2}^{h/2} r^2\rho v_{\f} v_r\,dz\right|_{r2} +\\[16pt]
\ds+\left.{1\over 4\pi}\int_{-h/2}^{h/2} r^2 B_rB_{\f}\,dz\right|_{r1}-\left.{1\over 4\pi}\int r^2 B_rB_{\f}\,dz\right|_{r2}+\\[16pt]
\ds + 2 \left(r\rho v_{\f}v_z-{1\over 4\pi}rB_{\f}B_{z}\right)_{S_{up}} r\,dr=0.
\label{angmom1}
\end{array}
\end{equation}
Integration across the disc is performed in the interval on z from $-{h\over 2}$ to ${h\over 2}$ correspondingly at the radiuses r1 and r2.
Integrations along $S_{up}$ and $S_d$ are equal to each other because the vector $d\bf S$ and the component of the velocity $v_z$ change sign simultaneously.
Therefore, we simply double the integration along the surface $S_{up}$. The terms containing magnetic field of the disc ${1\over 4\pi}\int_{-h/2}^{h/2} r^2 B_rB_{\f}\,dz$ are much less compared with the terms $\int_{-h/2}^{h/2} r^2\rho v_{\f} v_r\,dz$ provided that the condition
\begin{equation}
\rho V_k^2 \gg {1\over 4\pi} B^2
\label{condition}
\end{equation}
takes place. In other words the angular momentum flux is fully dominated by matter rather than the magnetic field in the disc. Under this condition Eq.(\ref{angmom1}) is reduced to the differential form as follows
\begin{equation}
\begin{array}{l}
\ds{\p\over r \p r} \left(r^2v_{\f} \int_{-h/2}^{h/2} \rho v_r\, dz\right)_{disc} \\[16pt]
\ds+2\left(r\rho v_{\f}v_z -{1\over 4\pi}rB_{\f}B_{z}\right)_{wind}=0.
\label{angmom22}
\end{array}
\end{equation}
The subscripts $disc$ and $winds$ denote the variables describing the disc and the wind at the base ( at the surface $S_{up}$).
According to this equation the angular momentum of the disc is carried out by the outflowing plasma and by the magnetic stresses in the outflow.

After similar manipulations it is easy to obtain that the equation for r-component of the momentum is as follows
\begin{equation}
\begin{array}{l}
\ds\left({1\over r}{\p \over \p r}\Big(r\int_{-h/2}^{h/2}\rho v_r^2 dz\Big)-{1\over r} \int _{-h/2}^{h/2}\rho v_{\f}^2\,dz \right)_{disc}\\[16pt]
\ds+ 2\left(\rho v_r v_z-{1\over 4\pi}B_rB_z\right)_{wind}=\left.
-\int_{-h/2}^{h/2}{\rho GM\over r^2}\,dz\right|_{disc}
\label{kepler}
\end{array}
\end{equation}
Here we neglect the thermal pressure, assuming that the disc is cold. This equation describes the balance of forces along radius in the disc. In the limit $v_r\to 0$ and under condition (\ref{condition})
 the Keplerian rotation takes place
\begin{equation}
{1\over r} \int_{-h/2}^{h/2} \rho v_{\f}^2\,dz={1\over r^2}\int_{-h/2}^{h/2} \rho\,dzGM.
\label{kepler2}
\end{equation}
Here we also neglected the term $\rho v_rv_z$ in the wind. Below we will see that this term is also equal to zero because in the case under consideration poloidal velocity of the wind equals to zero at the base.
Everywhere below we accept that the azimuthal velocity of the plasma in the disc $v_{\f}=V_k$.
The conservation law for the energy flux is as follows
\begin{equation}
\oint_S q_k\,dS_k=0
\end{equation}
Integration of the energy flux over the surface shown in Fig.~\ref{scheme} gives the following equation
\begin{equation}
\begin{array}{l}
\ds {1\over r}{\p\over \p r} \int_{-h/2}^{h/2} r\rho v_r\,dz \left({V_k^2\over 2}-{GM\over r}\right)_{disc} \\[16pt]
\ds+ 2\left(\rho v_z\left({v^2\over 2} -{GM\over R}\right)+ {1\over 4\pi}[E\times B]_z\right)_{wind}=0.
\label{energy1}
\end{array}
\end{equation}
The first term in this equation is the variation of the energy flux in the disc. Under condition (\ref{condition}) the energy flux of the disc consists only on the kinetic and gravitational energy of the matter. The magnetic field energy flux can be neglected. The energy from the disc is carried out by the wind. Radiation is neglected because the disc is cold.

The last equation necessary to connect the accretion with outflow is the matter conservation
\begin{equation}\label{matter}
{\p\over r\p r}\Big(r\int_{-h/2}^{h/2}\rho v_r\,dz\Big)_{disc}+\big(2 \rho v_z\big)_{wind}^{}=0.
\end{equation}

It is convenient to introduce the accretion rate in the disc
\begin{equation} \label{dotm}
\dot M = - 2\pi r\int_{-h/2}^{h/2} \rho v_r\,dz.
\end{equation}
In this case the last Eq.(\ref{matter}) takes the form
\begin{equation}
{\p \dot M\over \p r} -4\pi r\rho v_z=0.
\label{def1}
\end{equation}

If to insert this equation into (\ref{angmom22}), we obtain
\begin{equation}
\left.{\p\over \p r}\left( rV_k\dot M \right)\right|_{disc}-{\p \dot M\over \p r}\left.\left(r V_{k} - {rB_{\f}\over f}\right)\right|_{wind}=0.
\label{ang2}
\end{equation}
But the expression $\left(r V_{k} - {rB_{\f}\over f}\right)_{wind}$ equals to $L=r_A^2\Omega_k$ - the angular momentum per particle in the wind.
Differential equation for $\dot M$ takes the form
\begin{equation}
{\p \over \p r}(rV_k\dot M ) -{\p \dot M\over \p r}\,r_{\!A}^{}(r)^2 \Omega_k(r)=0,
\label{angmom2}
\end{equation}
where $r_A(r)$ is the Alfven radius of the force line rooted into disc at the point with radius $r$,
\begin{equation}
 \Omega_k(r) =\sqrt{GM\over r^3}
\end{equation}
is the Keplerian angular velocity of the disc at the radius $r$.
The solution of the equation (\ref{angmom2}) gives
\begin{equation}
\ln\!\left({\dot M\over \dot M_{\max}}\right)= -\int_r^{r_{\max}} {dr^2\over 4 (r_A^2(r)-r^2)},
\end{equation}
where $\dot M_{\max}$ is the accretion rate at the largest radius of the disc
$r_{\max}$. Let us consider the simplest case when $r_A(r)=\alpha r$, where $\alpha$ is some constant. The physical meaning of $\alpha$ is simple. This is
the Alfvenic radius of a force line expressed in the radius of the point at the disc where the force line is rooted. In this case the accretion rate in the disc varies with $r$ as follows
\begin{equation}
\dot M=\dot M_{\max} \left(r\over r_{\max}\right)^{{1\over 2(\alpha^2-1)}}.
\label{dotm2}
\end{equation}

The accretion rate in the case under consideration varies with $r$ according to power law. It is interesting that in the case $\alpha \gg 1 $ the power index appears close to zero.
This results into very interesting astrophysical implications which will be discussed below. Now we consider the role of Eq.(\ref{energy1}).

Using (\ref{def1}) it is easy to obtain from (\ref{energy1}) that
\begin{equation}
{\p \over \p r} \left.{\dot M V_k^2\over 2}\right|_{disc} +{\p \dot M\over \p r}\left({v^2\over 2}-{GM\over r} -{V_k B_{\f}\over f}\right)_{wind}=0.
\label{energy2}
\end{equation}
We use notations that ${1\over 4\pi} {[E\times B]_z\over \rho v_z}={\Omega_k r B_{\f}\over f}$ introduced in the previous section and the condition (\ref{kepler}).
The term $( {v^2\over 2}-{GM\over r} -{\Omega_k rB_{\f}\over f})_{wind}=E$. We are interested in the solutions which allows to particles to go to infinity from the disc. The necessary condition for this is
\begin{equation}
{v^2\over 2}-{GM\over r} -{\Omega_k rB_{\f}\over f}> 0.
\end{equation}
This condition means that the energy per particle is positive, what is necessary ( but not sufficient) to have positive $ v^2$ at large distance from the source.
The substitution of the explicit dependance of $\dot M$ into Eq.(\ref{energy2}) gives that
\begin{equation}
E=(2\alpha^2-3){GM\over 2r}
\label{particle}
\end{equation}
and the outflow of the wind at large distance from the central source is possible only under the condition $\alpha > \sqrt{3/2}$.

The integral relationships between the disc and the wind at the base allows us to define the velocity of plasma in the wind at the base.
Transformation of equation (\ref{energy2}) gives that on the disc surface
\begin{equation}
-{\p \dot M\over \p r} {V_k^2\over 2} + \dot M{\p \over \p r} {V_k^�2\over 2}+{\p \dot M\over \p r}\left({v^2\over 2} -{V_k B_{\f}\over f}\right)=0.
\label{energy3}
\end{equation}
Here we use the fact that $V_k^2 = {GM\over r}$.

Let us multiply equation (\ref{ang2}) by $\Omega_k$ and taking into account that $r\Omega_k =V_k$ transform the equation to the form
\begin{equation}
-{\p \dot M\over \p r} V_k^2 -\dot M {V_k^2\over r}-\dot M {\p \over \p r} {V_k^2\over 2} +{\p \dot M\over \p r}\left(V_kv_{\f}-{V_k B_{\f}\over f}\right)=0.
\end{equation}

If to take into account that ${V_k^2\over r}=-{\p V_k^2\over 2\p r}$, the equation above can be transformed as follows
\begin{equation}
-{\p \dot M\over \p r} V_k^2 +\dot M {\p \over \p r} {V_k^2\over 2} +{\p \dot M\over \p r}\left(V_kv_{\f}-{V_k B_{\f}\over f}\right)=0.
\label{angmom3}
\end{equation}

The substraction of Eq.(\ref{angmom3}) from (\ref{energy3}) gives
\begin{equation}
{\p \dot M\over \p r} {V_k^2\over 2}+{\p \dot M\over \p r}\left( {v^2\over 2}-V_kv_{\f}\right)=0.
\label{base1}
\end{equation}

The velocity of the wind at the disc surface $v^2=v_p^2+v_{\f}^2$. Therefore, after simple transformations we obtain from (\ref{base1}) that
\begin{equation}
v_p^2+(v_{\f}-V_k)^2=0.
\label{base2}
\end{equation}
It follows from (\ref{base2}) that at the base of the wind from the disc the poloidal velocity of the plasma equals to 0, and the azimuthal velocity equals to Keplerian
velocity of rotation. Physically this means that the wind fully corotates with the disc at the base.

\section{Selfsimilar solutions}

The accretion rate $\dot M$ appears power law function
of the disc radius if the Alfvenic radius is proportional
to the radius where the field line is rooted in the disc.
Therefore, it is interesting to look for selfsimilar
solutions of the problem

We are primarily concerned here with the plasma
dynamics immediately above the disc, at distances
$z$ much smaller than the radius of the disc $R_{\rm disc}$.
In the limit $z \ll R_{\rm disc}$, only two parameters with the
dimensions of length remain in the problem: $z$ and $r$.
Therefore, the solution will be self-similar in this limit
\citep{barenblatt}. Let us underscore an important feature of these
solutions: they describe flows only at small distances
from the disc, and are not applicable at distances
comparable to or exceeding the size of the disc.

\subsection{Selfsimilarity conditions}

The most comprehensive study of types of selfsimilar flows has
been investigated by the team from Athens University \citep{full}.
Here we will be interested in the selfsimilarity of the form proposed
initially by \cite{blandford} In this kind of selfsimilarity all
the variables depend on the coordinates in the form
\begin{equation}
 G(z,r)= r^{\delta} \t G\!\left({z\over r}\right),
\end{equation}
where $z,r$ are the cylindrical coordinates, and
$\delta$ is the selfsimilarity index.

The steady-state equations for an
ideal, cold plasma (with pressure P = 0) can be written as
\begin{equation}\label{e1}
\rho({\bf v}\nabla){\bf v}=-{1\over 8\pi}\nabla\,{\bf B}^2+{1\over 4\pi}({\bf B}\nabla){\bf B}
-\rho\,\frac{GM\bf R}{R^3}\,,
\end{equation}

This equation defines the structure of the wind.
According to the selfsimilarity assumption all the variables in these equations can be presented as
\begin{equation}\label{au}
\begin{array}{rcl}
{\bf v}(r,z,\phi)&=&r^{-\delta_v}_{}\t{\bf v}(z/r,\phi)\,,\\[4pt]
\rho(r,z)&=&r^{-\delta_{\rho}}_{}\t\rho(z/r)\,,\\[4pt]
{\bf B}(r,z,\phi)&=&r^{-\delta_B}_{}\t{\bf B}(z/r,\phi)\,.
\end{array}
\end{equation}

This representation of the variables says that they are scaled as the power law of
$r$ and all functions with $\t{}$ depend on the angle $\xi$ defined as $\tan\xi = z/r$.

The superscripts $\delta_v$, $\delta_{\rho}$, and $\delta_B$ are determined from
the following conditions. Substituting (\ref{au}) into (\ref{e1})
leads to the equations
\begin{equation}\label{ind1}
2\delta_B-\delta_\rho=2\delta_v=1\,.
\end{equation}
It follows from these equations that

\begin{equation}\label{auto}
\begin{array}{rcl}
\ds{\bf v}(r,z,\phi)&=&r^{-1/2}_{}\,\t{\bf v}(z/r,\phi)\,,\\[4pt]
\ds\rho(r,z)&\,=\,&r^{-\delta}_{}\t\rho(z/r)\,,\\[4pt]
\ds{\bf B}(r,z,\phi)&=&r^{-{(1+\delta)\over 2}}_{}\,\t{\bf B}(z/r,\phi)\,.
\end{array}
\end{equation}
Only one index $\delta$ defines the family of the solutions. In our previous work this
index has been fixed by the condition (\ref{frost}) and (\ref{varphi}). Formally it
follows from these conditions that $vB \sim R^{-1}$ and therefore
$\delta = 0$. However, it is worth to pay attention that in the relationship between
$E_{\f}$, $B$ and $v$ only the component of the velocity $v_{r}$ perpendicular to the
poloidal field line comes in. But this component does not play any role in the dynamics of the disc or the
wind because we consider the case $v_{r} \ll V_k$. This component of the velocity is present neither in the
wind equations or in the equations connecting the disc and the wind. Therefore, we can relax the condition of the selfsimilarity and consider a more general case when
the component of the velocity $v_{r}$ in the disc does not follow to the selfsimilarity prescriptions (\ref{au}). The index $\delta$ remains
free parameter of the problem in this case.

Let us now consider how the index $\delta$ is connected with the index $\alpha$ in the power law dependence of the accretion rate.
The position of the Alfven point is defined by the condition.
\begin{equation}
v^2={B_p^2\over 4\pi\rho}.
\end{equation}
 Substitution of the conditions (\ref{au}) in this relationship gives that at the Alfven point the ratio $z_A{}/r_A^{}$ is constant
for all field lines. All the filed lines in the selfsimilar solutions differ by scaling of the coordinates. The shape of a field line is described by the equations
\begin{equation}\label{line}\left.
\begin{array}{lcl}
z &=& r_0^{}\,g(\xi) \tan(\xi)\,,\\[4pt]
r &=&r_0^{}\,g(\xi)\,,\\
\end{array}
\right\}
\end{equation}
where $r_0^{}$ is the cylindrical radius of the field line at the base.
It follows from these relationships that at the Alfven point the ratio $r_A^{}/r_0^{}=\alpha$ is constant for all the field lines. This means that in the selfsimilar solutions under the consideration the accretion rate varies with radius according to (\ref{dotm}).

Substitution of (\ref{dotm}) into (\ref{def1}) taking into account (\ref{auto}) gives the following relationship between $\alpha$ and $\delta$
\begin{equation}
\delta={3\alpha^2-4\over 2\,(\alpha^2-1)}.
\end{equation}
 $\alpha$ can change from ${3\over 2}$ up to $\infty$. It is interesting that at the same time the index of selfsimilarity varies only in the limits between ${1\over 2}$ and ${3\over 2}$.

It is interesting also to estimate the dependence of radial velocity of the matter in the disc on radius. Taking into account that the toroidal electric field depends on $r$ in the disc as $E_{\f} \sim r^{-1}$ and the magnetic field varies as $B \sim r^{-{1+\delta\over 2}}$. The frozen-in condition gives that the radial velocity depends on $r$ as $v_r \sim r^{{\delta-1\over 2}}$. The radial velocity falls down with $r$ or remains constant at $\delta ={1\over 2}$.

\subsection{Dimensionless variables}

Is convenient to consider the flow in the dimensionless variables. In the selfsimilar solution all the geometrical variables can be scaled in the units of $r_0^{}$, the radius at the disc where a selected field line is rooted. The magnetic and electric fields are dimensionlessed by the value $B_0={\psi\over \pi\,r^2}$, where $\psi$ is the magnetic flux through the disc with radius $r_0^{}$. Kepler velocity $V_{k}$ at the radius $r_0^{}$ is used to dimensionless the velocities. These variables provides us a characteristic density
\begin{equation}
 \rho_0= {B_0^2\over 4\pi V_k^2},
\end{equation}
 which is used to dimensionless the density of the plasma. It follows from the assumption ${B^2\over 4\pi \rho_d} \ll V_K^2$ that the
dimensionless density of plasma in the disc is much greater than~1.

\section{Results}

The technic of solution of the problems of selfsimilar MHD outflows has been developed in many papers starting with \cite{blandford}.
The most comprehensive study of the selfsimilar outflows is presented by \cite{method}. In these papers the methodology of the solution is also presented in details.

The most difficult task at the solution of the MHD problems of outflows is the selection of the unique solution regular at the critical points: slow magnetosonic, Alfvenic and fast magnetosonic critical points. In our problem the slow magnetosonic critical point is absent because the outflow is cold. It is necessary to find a solution regular at the Alfvenic and fast magnetosonic critical points.

Typically the selfsimilar solutions are regularized at the Alfvenic point. It is sufficient to regularize the solution at the Alfvenic point to specify the selfsimilar solution \citep{blandford}. The group of Athens university was succeeded to obtain the selfsimilar
solution regular in both Alfvenic and fast magnetosonic critical points \citep{fastmode}.

\begin{figure}
\centerline{\includegraphics[width=0.5\textwidth,angle=-90.0]{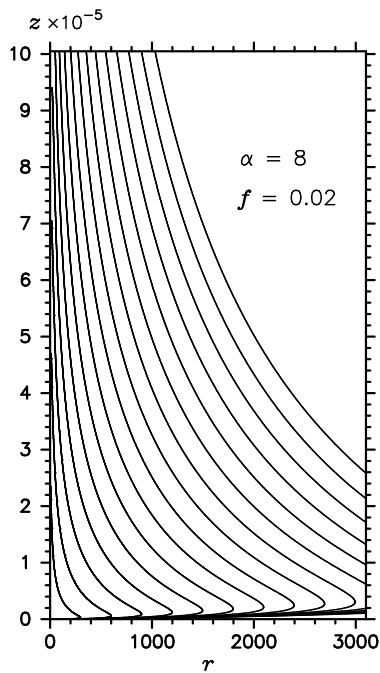}}
\caption{The filed lines for the solution corresponding to $\alpha=8$, $f=0.02$.}
\label{solution}
\end{figure}

In our work we follow to the conventional approach and look for the solutions regular at the Alfvenic point only.
For astrophysical implications the solutions with large $\alpha$ are the most interesting. We present here the results for $\alpha = 8$ and $f=0.02$. Field lines are shown in Fig.~\ref{solution}. The field lines start from the disc under the angle $21^{\circ}$. Therefore the  \cite{blandford} condition of outflow of matter from the disc is fulfilled. Interesting feature of this solution is that initially the field lines of the solution diverge. However, at some distance from the disc they start to recollimate forming the jet-like flow. The scale along z axis is strongly reduced in Fig.~\ref{solution}. The solution reaches distances greater than million times of the initial radius. Nevertheless, the flow does not cross the fast mode critical point.

\begin{figure}
\centerline{\includegraphics[width=0.5\textwidth,angle=-90.0]{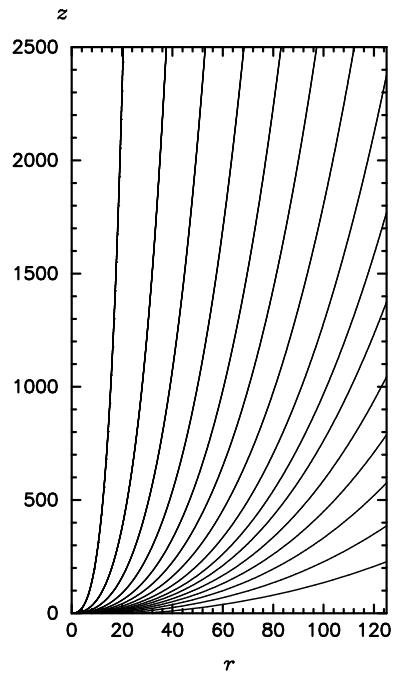}}
\caption{The filed lines for the solution corresponding to $\alpha=5$, $f=0.1$.}
\label{sol4}
\end{figure}

This behavior is typical for large $\alpha$. For smaller $\alpha$ the field lines looks like in the solutions obtained by Blandford \& Payne (1982).
The field lines for $\alpha=5$ are shown in Fig.~\ref{sol4}. They start from the disc under the angle $16^{\circ}$.

Such dependance on $\alpha$ of the solution is understandable. The larger $\alpha$ the larger the magnetic field or smaller the mass flux density along a field line. In any way this results into stronger affect of the magnetic field on the flow. As the result, at larger $\alpha$ the effect of the magnetic collimation appears stronger.

\begin{figure}
\centerline{\includegraphics[width=0.33\textwidth,angle=-90.0]{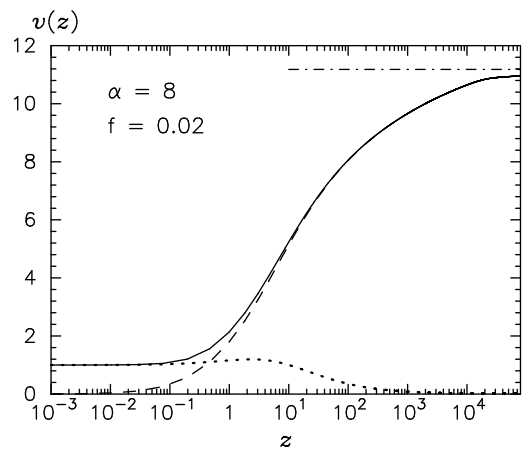}}
\caption{Variation of velocity with $z$ for $\alpha=8$, $f=0.02$. Solid line - full velocity, dashed line - poloidal velocity, dotted line - azimuthal velocity. Horizontal dashed-dotted line shows the upper limit on the velocity following from Eq.(\ref{particle}) }
\label{vel}
\end{figure}

In Fig.~\ref{vel} we show the variation of the plasma velocity along a field line with $z$. The growth of the velocity occurs at $z$ comparable with $r$. It is interesting that the terminal velocity more than 10 times exceeds the Keplerian velocity.

\begin{figure}
\centerline{\includegraphics[width=0.33\textwidth,angle=-90.0]{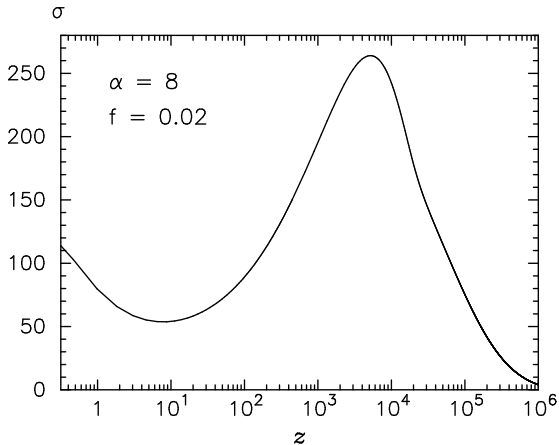}}
\caption{Dependance of the magnetization of plasma on $z$. }
\label{sigma}
\end{figure}

Surprisingly, the transformation of the Pointing flux into kinetic energy flux in the appeared very efficient. The ratio of the Pointing flux over the kinetic energy flux in the winds conventionally characterized by the $\sigma$ - parameter. The behavior of this parameter along a field line is shown in Fig.~\ref{sigma}. Initially the wind is strongly magnetically dominated, but at high distances the energy is fully transformed into bulk motion kinetic energy.

\section{Discussion}
\subsection{Energy budget}

According to our solution the accretion rate $\dot M$ goes to zero at the gravitation centre. This means that no matter falls onto the centre. All the accreted matter is ejected. Nevertheless, in the solutions with $\alpha^2 > 1.5$, energy per particles in the wind is positive and therefore the wind carries positive energy. This produces impression that conservation of the energy is violated.

Apparent energy conservation violation certainly deserves of a detailed consideration. First of all we have to point out that the violation of the energy conservation in our solution is impossible, because the energy conservation law has been used explicitly in the solution. In the wind the energy is conserved by imposing that $E$ from (\ref{energy}) is constant along any field line. At the same time the dynamics of the wind is connected with the dynamics of the disc by the conservation laws: mass, angular momentum and energy.

\begin{figure}
\centerline{\includegraphics[width=0.4\textwidth,angle=0.0]{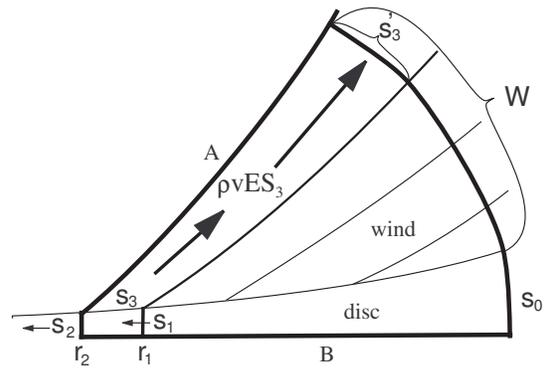}}
\caption{The integration of the energy flux over the surface surrounding the disc. The surface is shown by thick lines. The gas inflow into the
disc at it outer edge with the energy per particle $E=0$ gives zero contribution into the total budged. $E >0$ everywhere in the wind. However, the matter
crossing the surface $S_2$ carries flux of negative energy which totally compensate the flux of energy of the wind.}
\label{problem4}
\end{figure}

The conservation of the energy in the disc - wind system follows from equation (\ref{energy2}). Let us select some fragment of the disc, surrounded by the surfaces $S_1$, $S_2$ and $S_3$ as it is shown in Fig.~\ref{problem4}. According to (\ref{energy2}) the energetic budget of this fragment is as follows

\begin{equation}
\dot M(r_1)\,{GM\over 2r_1}- \dot M(r_2)\,{GM\over 2r_2} + (E\rho v_z S_3)_{wind}=0.
\label{energy33}
\end{equation}

This relationship is valid for the disc and the wind at the base. Nevertheless, the values $E$ and $\rho v_p S_3'$ are conserved along filed lines. Here $v_p$ is the poloidal velocity of the wind and $S_3'$ is the square between two neighbour field lines. Therefore, in Eq.(\ref{energy33}) we can replace $E\rho v_z S_3$ on the value
$E\rho v_p S_3'$ which correspond to any point located arbitrary far from the disc. Equation (\ref{energy33}) can be summed over all similar fragments of the disc and attached wind starting with some value of $r_{2}$ up to the outer edge of the disc. After that only two terms along surfaces $S_2$ and $W$ survive. Integrals along surfaces $A$ and $B$ equal to zero because there is no matter flux through them. Integral along the surface $S_0$ located at the outer edge the disc
equals to zero because $E=0$ on this surface. Thus we have
\begin{equation}
-\dot M(r_2)\,{GM\over 2r_2} +\int_W E\rho v_p\, dS=0.
\label{energy4}
\end{equation}

The value $\int_W E\rho v_p\,dS =L_{W}$ is the flux of the energy in the wind which is positive in the case under consideration. It is exactly balanced by the flux of negative energy which flow to the gravitation centre through the surface $S_2$. Thus, total flux through any surface containing the disc and the wind as it is shown in Fig.~\ref{problem4} equals to zero. This means that all the energy flux in the wind is the gravitational energy released at the accretion of the mass remaining in the disc.

The solution of the paradox with the apparent violation of the energy conservation directly follows from Eq.(\ref{energy4}). The mass flux onto gravitation centre $\dot M(r)$ goes to zero at the centre indeed. Nevertheless, the negative energy flux which is the product $ -\dot M(r) {GM\over 2r}$ diverges at the centre for $\alpha^2 > 1.5$. This flux of negative energy exactly equals the positive flux of the energy of the wind.

The budget consideration demonstrates that the solution describes the process of transfer of the energy of matter infalling onto gravitating centre to the matter ejected from the disc. The magnetic field provides the mechanism for this transfer.

\begin{figure}
\centerline{\includegraphics[width=0.35\textwidth,angle=0.0]{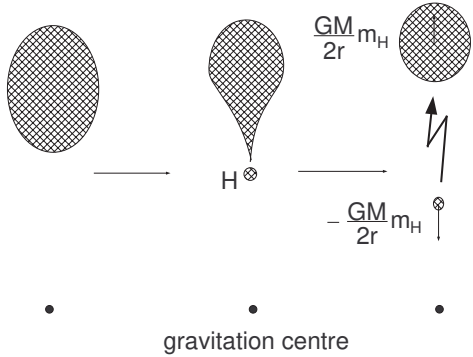}}
\caption{ Illustration of energy conservation at the ejection-accretion processes. The matter shown by shadowed region falls initially with zero energy. On the way to the gravitation centre one hydrogen atom (infinitesimally small part) splits from the matter and falls separately. Magnetic field provides mechanism of the energy transfer from the atom to the rest of mass which is ejected with infinitely large energy while the atom approaches the centre. }
\label{example}
\end{figure}

This paradox physically is illustrated by Fig.~\ref{example}. In some sense it is analog of well known Penrose process of energy extraction from a black hole (Penrose, 1969).
Let us imagine that some amount of matter is dropped onto a gravitating centre with zero initial energy. On the way to the gravitating centre infinitesimally small amount of matter (say one hydrogen atom) splits from the matter and falls down onto the centre separately. The energy of this atom is $E_H^{}=-{GM\over 2\,r}\,m_H^{}$ and goes to minus infinity with $r\to0$. The energy equal to $-E_H^{}$ is transferred to the remaining mass by means of the magnetic field. Due to this the remaining mass will be ejected from the gravitation centre with the energy $-E_H$. While the hydrogen atom approach to the gravitation centre, $E_H$ goes to $-\infty$. This means that the remaining mass will be ejected with infinite positive energy.

This way infinitesimally small amount of mass falling onto the gravitating centre results into ejection of finite mass with infinite energy. Of course this is the result of gravitational singularity. In application to the real astrophysical objects it is necessary to take into account that some minimal possible radius always exists. In the case of stars (neutron or protostars) this is the radius of the stars, in the case of black holes this is Schwarzschild radius. If so,
the accretion disc has some inner radius $r_{in}$ and the relationship between the energy flux in the wind and the accretion rate is as follows

\begin{equation}
L_{W}=\dot M\, {GM\over 2r_{in}}= \dot M\, {V_k^2\over 2r_{in}}
\label{wind}
\end{equation}

The relationship (\ref{particle}) gives $E_{particle}$ in the wind. It is useful to present this energy in the form

\begin{equation}
E_{particle} =(2\alpha^2-3)\, {mV_k^2\over 2}.
\label{parten}
\end{equation}

There is rather conventional opinion (Livio) that the energy of the particles in the wind from the disc should be of the order of Keplerian velocity.
It follows from (\ref{parten}) that the particle energy can remarkably exceed the Keplerian energy provided that $\alpha^2$ well exceeds 1.5.

\subsection{Efficiency.}

The efficiency of the outflow formation by a system can be defined as the ratio of the kinetic luminosity of the outflow ( jet plus wind) over the energy released at the accretion. According to Eq.(\ref{wind}) the efficiency of the system under the consideration is equal to 1. Almost $100\%$ of accretion energy is transformed into the energy of the outflow.

At the estimate of the efficiency of real objects bolometric luminosity is used as the estimate of the energy released at the accretion. This estimate follows from \cite{shakura} solution. Therefore, the efficiency $\eta$ equals to
\begin{equation}
\eta={L_{W}\over L_{bol}}.
\end{equation}

In our solution bolometric luminosity equals to zero because all the energy goes to the energy of the wind and no processes which transform the energy of plasma into the thermal energy. Formally, for our solution $\eta=\infty$. However, in real objects dissipative processes, viscosity and finite electric conductivity, always takes place and some fraction of energy will go into the radiation of the system. But the energetics of this radiation does not already directly connected with the energetics of accretion like in the \cite{shakurasunyaev} solution. Therefore, we can expect that for some astrophysical objects $\eta$ could be very high. This conclusion naturally explains the ratio of the kinetic luminosity on the jet from M87 over the bolometric luminosity of this object which archives the value as much as 100.

\subsection{Jet formation.}

The case when $\alpha$ well exceeds $\sqrt{3/2}$ is especially interesting for astrophysical implications. Such a wind from the accretion disc provides the rate of accretion practically constant along the radius of the disc. Indeed, the disc wind with $\alpha =5$ gives that the rate of accretion
varies as $\dot M \propto r^{1/48}$, while for $\alpha =8$ the dependance becomes $\dot M\propto r^{1/126}$. The accretion rate appears almost constant in the largest part of the disc and sharply reduces at the centre. This means that the outflow rate is weak from the disc and the most dense and fast outflow occurs from the very central part of the accretion disc. The matter flux density varies as
\begin{equation}
\rho v={{\dot M_{\max}}\over{2\,(\alpha^2-1)\,r^2}}\left(r\over r_{\max}\right)^{{1\over 2(\alpha^2-1)}}
\end{equation}
It follows from this equation that increase of $\alpha$ reduces the mass flux from the disc. This is natural, because increase of $\alpha$ means increase of the angular momentum carried out by every particle of the wind. Therefore, the same angular momentum can be carried out by less amount of the wind.

Obtained form of the accretion rate explains rather puzzling properties of the jets. Jets always ar formed at the central part of the system producing impression that some way the central source is responsible for the jet production. In the solution we see that practically all the outflow concentrates at the very central part of the disc.

\begin{figure}
\centerline{\includegraphics[width=0.45\textwidth,angle=0.0]{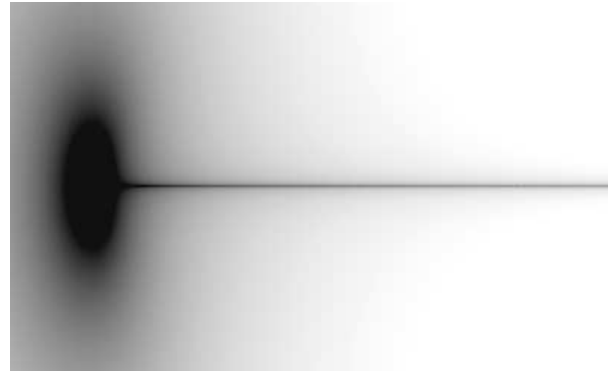}}
\caption{Distribution of column density along line of sight for the solution with $\alpha=8$, $f=0.02$. }
\label{jet}
\end{figure}

To illustrate this property of the solution we have calculated for the solution with $\alpha=8$ the distribution of column density along line of sight.
The viewing angle of the disc is $42^{\circ}$. Integration of the density near the disc was limited by the points on the field lines which are located at the Alfvenic radiuses. Therefore the disc appeared visible only at the central part. Otherwise integrals diverges at the disc. The result of calculation is presented in Fig.~\ref{jet}. This figure well illustrates that the obtained solution naturally explains why the jets are formed at the very central part of the accretion disc.

\section{Conclusion.}

The selfconsistent solution of the problem of the disc accretion and plasma outflow from the disc shows that such puzzling properties observed in real astrophysical objects as high energetic efficiency of jets, launching of jets from very central part of the accretion discs and high velocities of outflows which in some objects well exceeds the Keplerian velocities, are naturally explained in frameworks of general approach assuming that the largest part of the angular momentum from the disc is carried out by the wind rather than due to viscous stresses. Although a significant work should still be done to explore the astrophysical implications of the obtained solution to specific astrophysical objects and generalization of the solution for relativistic objects, we believe that this approach is rather promising for explanation of properties of jets from objects of different nature, starting with jets from protostars up to jets from AGN.

\section{Acknowledgments}

S Bogovalov and S. Kelner are grateful to the High Energy Astrophysics Group of Max-Planck institute for warm hospitality. The work of S.~Kelner has been performed during his long-term stay in the Max-Planck Institute f\"ur Kernphysik. We are especially grateful to Felix Aharonian who forced us to perform this study.

\end{document}